# The Effect of Ru substitution for Ni on the superconductivity in $MgCNi_{3-x}Ru_x$

**T. Klimczuk[1,2], V. Gupta[1], G. Lawes[3], A.P. Ramirez[3] and R.J. Cava[1]**

[1]Department of Chemistry, Princeton University, Princeton NJ 08544,
[2]Faculty of Applied Physics and Mathematics, Gdansk University of Technology,
Narutowicza 11/12, 80-952 Gdansk, Poland,
[3]Los Alamos National Laboratory Los Alamos, New Mexico 87544,

The superconductor $MgCNi_3$ has been chemically doped by partial substitution of Ru for Ni in the solid solution $MgCNi_{3-x}Ru_x$ for $0<x<0.5$. Magnetic and specific heat measurements show that the Sommerfeld parameter ($\gamma_{exp}$) and $T_C$ decrease immediately on Ru substitution, but that a $T_C$ above 2K is maintained even for a relatively large decrease in $\gamma_{exp}$. Ferromagnetism is not observed to develop through Ru substitution, and the normal state magnetic susceptibility is suppressed.

## Introduction

For a handful of materials, the evolution from superconductivity to ferromagnetism as chemical, structural or electronic characteristics are changed is the subject of considerable current research[1-3]. The 7K superconductor $MgCNi_3$ displays none of the extreme sensitivities to impurities and disorder that the mainstream of such compounds display[4], but nonetheless its very large Ni content, and unusual characteristics of its calculated electronic structure, notably a very large, narrow peak in the density of states at the Fermi Energy, have suggested that it may have an exotic mechanism for superconductivity[5-12]. Experimental characterization of the superconducting and normal state properties have so far revealed some properties that appear to be entirely conventional in character[13], and others that are highly unusual[14,15].

Analysis of the calculated electronic structure has suggested that $MgCNi_3$ is electronically close to a ferromagnetic instability, and that a change in electron count, as little as - 0.1 electrons per formula unit in some analyses, should put the Fermi level near the peak in the expected electronic density of states, leading to a ferromagnetism[12,16]. This has motivated a series of studies to decrease the electron count $MgCNi_3$ by chemical alteration to uncover the incipient ferromagnetic state: by inducing carbon deficiency[17,18], and by partial substitution of Co, Fe or Mn for Ni[19-21]. The substitution of $3d$ elements for Ni suppresses the superconductivity to below 2K very quickly, at levels of 1% substitution or less, but, though an increase in magnetic susceptibility is found, the superconductivity does not to evolve into ferromagnetism. The fact that $T_C$ is suppressed so quickly suggests that the $3d$ element substitutions have suppressed $T_C$ by magnetic pair breaking and not through a band structure (i.e. electron count) effect.

Here we report the effect of Ru substitution on the superconductivity in $MgCNi_3$, with the goal of inducing primarily an electron count effect by substitution of an electron deficient element that is not likely to act as a magnetic pair breaker due to its $4d$ character. Characterization of the materials prepared indicates a decrease of magnetic susceptibility with increasing Ru content, the opposite of what is observed for the Co and Mn substitutions, indicating that Ru displays no magnetic moment. We show that the superconductivity is suppressed much more slowly than is observed in the $3d$ element substitution cases, though again, no ferromagnetism is revealed.

## Experimental

A series of 0.5g samples with compositions $Mg_{1.2}C_{1.5}Ni_{3-x}Ru_x$ (x=0, 0.005, 0.01, 0.02, 0.033, 0.066, 0.1, 0.2, 0.3, 0.4, 0.5, and 0.6) were synthesized. The starting materials were bright Mg flakes (99+% Aldrich Chemical), fine Ni powder (99.9% Johnson Matthey and Alpha Aesar), glassy carbon spherical powder (Alfa AESAR), and Ruthenium powder (99.95% Alpha Aesar). Previous studies on $MgCNi_3$ indicated a need to incorporate excess magnesium and carbon in order to obtain optimal carbon content[4,17]. The excess Mg is vaporized in the course of the reaction. The resulting material is stoichiometric $MgCNi_3$ plus a small proportion of elemental carbon[17]. After thorough mixing, the starting materials were pressed into pellets, wrapped in Tantalum foil, placed on an $Al_2O_3$ boat, and fired in a quartz tube furnace under a 95% Ar, 5% $H_2$ atmosphere. The initial furnace treatment began with a half hour at 600°C, followed by 1 hr at 900 C. After cooling, the samples were reground, pressed into pellets, and placed back in the furnace under identical conditions at 900°C. The latter step was repeated two additional times. Following the heat treatment, the samples were analyzed through powder X-ray diffraction (Cu Kα radiation).

Zero field cooling DC ($H_{DC}$=20Oe) and AC ($H_{DC}$=5 Oe, $H_{AC}$=3 Oe, f=10 kHz) magnetizations were measured in the range of 1.8K to 8K (PPMS – Quantum Design). The specific heat measurements were done using a standard relaxation technique with a commercial calorimeter (Quantum Design PPMS). In order to ensure good thermal contact for the specific heat measurements, the samples were mixed with fine silver powder in a 1:1 ratio and cold sintered into a hard disk. The heat capacity of the silver powder was measured separately and subtracted.

## Results

The X-ray diffraction measurements indicated that the crystallographic cell parameter for the $MgCNi_{3-x}Ru_x$ phase increases systematically with increasing Ru content. The extended powder pattern for $MgCNi_3$ is shown in the inset to figure 1. Very small amounts of MgO were observed in some samples. The MgO present does not affect the stoichiometry of the $MgCNi_3$ phase, and neither does any elemental C present.[17] The main panel of figure 1 shows the shifting of the (332) peak for the six representative doped samples with stoichiometry $MgCNi_{3-x}Ru_x$ (x=0, 0.1, 0.2, 0.3, 0.4, and 0.5). The distinct $\alpha_1$-$\alpha_2$ splitting confirms the high sample quality – such splitting would not be visible for poorly crystallized or compositionally inhomogeneous materials. Figure 2 indicates a linear relationship between the Ruthenium concentration and the cubic cell parameter. As the Ru doping is increased from x=0 to x=0.5, the lattice expands from a=3.809(1) Å to 3.851(1) Å, as derived from least-squares fits to 9 X-ray reflections between 20 and 90 degrees 2θ. The x=0.6 sample is multiple phase, indicating that the limit of solubility is between 0.5 and 0.6.

The reported $T_C$s for pure bulk $MgCNi_3$ vary between approximately 6 and 7.5K[4,18,21]. Such differences may be due to the carbon content of the samples, but in the initial preparation of the $MgCNi_3$ samples for this study we observed that under the same synthetic conditions the $T_C$ observed was dependent on the starting materials employed. As figure 3 shows, when Johnson Matthey Puratonic Nickel Powder and Johnson Matthey Nickel Powder – Low Carbon powders were used, the resulting $MgCNi_3$ samples gave a superconducting transition temperature ($T_C$ onset) around the reported value of 7K. However, when Alfa Aesar Nickel Powder was employed, the transition temperature dropped to 5.7K. We presume that a minor magnetic impurity in the Alfa Aesar Nickel powder is responsible for the reduction in the transition temperature. As a result of this finding, all samples used for the purposes of this paper were synthesized using Johnson Mathey Puratronic Ni.

Figure 4 shows, for representative Ru concentrations, zero-field cooling DC magnetization data for the $MgCNi_{3-x}Ru_x$ samples. $T_C$ decreases systematically with increasing Ru content. The transitions remain bulk in character, but as $T_C$ approaches the minimum available temperature of the measurement, smaller diamagnetism and a smaller fraction of the transition is observed as less of the transition is accessible in the temperature range of measurement. In order to measure the transition temperatures most precisely, zero field AC susceptibilities were measured. The AC data for the same samples are shown in figure 5. The $T_C$s determined from the AC data are used in the analysis that follows. As seen in figure 8, the superconducting transition temperature decreases systematically with increasing Ru content. The rate of the decrease in $T_C$ is very much slower than is observed for the electron-deficient substitutions from the $3d$ series. Tc decreases more slowly at the larger Ru concentrations. Because measurements could not be made below 1.8K, we do not know whether x=0.5 displays superconductivity at a lower temperature.

Unreacted ferromagnetic Ni metal, always present in very small (fractional percentage) amounts in $MgCNi_3$ powder preparations (no single crystals have yet been reported), complicates measurement of the normal state magnetic susceptibility.

This contribution to the susceptibility must be subtracted through the analysis of M vs. H curves. Figure 7 illustrates the field dependence of the magnetization for the six representative samples of $MgCNi_{3-x}Ru_x$ measured at T=10K. Initially, the magnetization for all six samples increases rapidly, until $\mu_0H$ equals approximately 2T: the initial rapid increase in the magnetization is due to the presence of Ni metal impurity, and is present in small amounts to the highest temperature of our measurements (300K). Therefore, to approximate the intrinsic susceptibility of the $MgCNi_{3-x}Ru_x$ compounds, the difference in magnetization ($\Delta M$) between applied fields of 4T and 2T was employed at each temperature to estimate the susceptibility ($\chi=\Delta M/\Delta H$) for all samples. The maximum amount of Ni present, estimated from the saturation magnetization of Ni metal, is 0.1%.

The susceptibility ($\Delta M/\Delta H$) derived in such a fashion for representative samples between x=0 and x=0.5 is shown in figure 8, for temperatures between 5K and 300K. The susceptibilities increase with decreasing temperature, as previously observed for $MgCNi_3$[19,20]. Both the magnitude of the susceptibility and its increase with decreasing temperature are suppressed by increasing Ru substitution. The susceptibility for x=0.5 is essentially temperature independent paramagnetism. The Ru concentration dependence of the estimated susceptibility at 10 K is presented in figure 9. There is a substantial decrease in the susceptibility at 10K, even for the smallest Ru doping levels.

Figure 10 presents the specific heat at low temperatures for representative Ru concentrations. The specific heat above $T_c$ consists of the linear electronic contribution ($\gamma_{exp}T$) and the cubic lattice contribution ($\beta T^3$). Plotting C/T as a function of $T^2$ allows us to extract values for $\gamma_{exp}$ from the intercept of the straight line that is the best fit to the data from just above $T_C$ to $T^2=125$ $K^2$. Some deviation from linearity visible for the x=0.5 sample at the lowest temperatures suggests that there may be some kind of phase transition for T < 2K for this material. The composition dependence of $\gamma_{exp}$ is shown in figure 11. For our undoped $MgCNi_3$, the Sommerfeld parameter $\gamma_{exp}$ is 31±2 mJ/mol $K^2$, which is in good agreement with values previously reported $(29 - 33.6$ mol/mol $K^2)$[4,13,14,16,19,20]. There is a dramatic decrease in $\gamma_{exp}$ at very low Ru concentrations, from 31 mJ/mol $K^2$ at x=0 to 24 mJ/mol $K^2$ by x=0.05. $\gamma_{exp}$ continues to decrease with increasing Ru concentration, but decreases at a much lower rate between x=0.1 and x=0.5 then it does between x=0 and x=0.1.

## Discussion and Conclusions

Band structure calculations on $MgCNi_3$ universally agree that the density of states at the Fermi level should first increase and then decrease on decreasing the electron count of $MgCNi_3$[5-12]. Figure 11 shows, however, that the Sommerfeld parameter $\gamma_{exp}$ decreases over the whole range of electron deficiency afforded by Ru substitution, and to the extent that $\gamma_{exp}$ is a representation of the electronic density of states, the current results suggest that that does not occur for electron count decrease induced by Ru substitution.

Figure 12 illustrates the variation of $T_C$ with Sommerfeld parameter, $\gamma_{exp}$. For low levels of doping, (the highest $\gamma_{exp}$ values) $T_C$ changes very little with substantial changes in $\gamma_{exp}$: a decrease in $\gamma_{exp}$ of about 20%, from 31 to 25 mJ/mol $K^2$, changes $T_C$ by less than 1K. The relative insensitivity of $T_C$ to electron count, structural disorder, and $\gamma_{exp}$ clearly distinguishes $MgCNi_3$ as being substantially different from superconductors where ferromagnetism and superconductivity have been clearly linked. Finally, figure 13 presents the magnetic susceptibility, $\Delta M/\Delta H$, plotted against $\gamma_{exp}$ (data taken from figures 9 and 11). The figure shows that the magnetic susceptibility increases as $\gamma_{exp}$ increases, in agreement with the expected behavior. It further shows that within experimental error the ratio of $\chi$ ($\Delta M/\Delta H$) to $\gamma_{exp}$ is estimated to be 1.11±0.05 $\left( R = \dfrac{\pi^2 k_B^2}{3\mu_B^2} \dfrac{\chi}{\gamma_{exp}} \right)$ for the whole range of Ru substitution. The present value is close to the 1.2 reported for the undoped material by Hayward[20].

Our characterization of the superconductivity in the solid solution series $MgCNi_{3-x}Ru_x$ has shown that the effect of the electron deficient $4d$ metal substitution for Ni is substantially different from what is observed in the cases of $3d$ element substitution. $T_C$ is relatively robust across the $4d$ substitution solid solution, suggesting that the strong suppression of $T_C$ for the $3d$ substitutions represents magnetic pair breaking. There may be mitigating effects of disorder, and differences in coupling strength across the series, but the observed behavior of $\gamma_{exp}$ on Ru substitution is not consistent with the presence of a sharp peak in the electronic density of states at electron counts just less than that of $MgCNi_3$, consistent with XPS, XAS and UPS spectroscopy studies that have suggested that the peak in the DOS is strongly renormalized from what is seen in electronic structure calculations[9,23]. Considering that Cu doping (which adds electrons) also suppresses $T_C$ without the development of ferromagnetism, the current results suggest that ferromagnetism in $MgCNi_3$ is not easily induced by chemical substitution.

## Acknowledgements

This work was supported by the US Department of Energy, grant DE-FG02-98-ER45706 and partially by the US National Science Foundation, grants DMR 0244254 and 0213706. Work at Los Alamos National Laboratory was supported by LDRD.

The work done by Tomasz Klimczuk was graciously supported by The Foundation for Polish Science (Foreign Postdoc Fellowship Grant).

**Figure captions:**

**Fig.1** Powder X-ray diffraction data for $MgCNi_{3-x}Ru_x$ ($CuK_\alpha$). Inset: full angular scan. Main panel, (332) peak showing $\alpha_1$ - $\alpha_2$ split

**Fig. 2** (Color online) Cubic cell parameter vs. x in $MgCNi_{3-x}Ru_x$. Inset shows the crystallographic structure

**Fig. 3** Magnetic characterization of the superconducting transition for three samples of $MgCNi_3$ prepared from different Ni starting material

**Fig. 4** DC magnetization characterization of the superconducting transition in representative $MgCNi_{3-x}Ru_x$ samples

**Fig. 5** AC magnetization characterization of the superconducting transition in representative $MgCNi_{3-x}Ru_x$ samples

**Fig. 6** Superconducting critical temperature ($T_C$) in $MgCNi_{3-x}Ru_x$ as a function of Ru doping (x)

**Fig. 7** Field dependence of the magnetization at 10K for $MgCNi_{3-x}Ru_x$ used to extract the effect of ferromagnetic Ni impurity

**Fig. 8** Magnetic susceptibility $\chi$ vs. temperature in representative samples of $MgCNi_{3-x}Ru_x$. $\chi = \Delta M/\Delta H$ determined as shown in figure 7

**Fig. 9** Magnetic susceptibility $\chi$ at 10K as a function of x in $MgCNi_{3-x}Ru_x$. $\chi = \Delta M/\Delta H$ determined as shown in figure 7

**Fig. 10** Low temperature specific heat characterization of representative $MgCNi_{3-x}Ru_x$ samples

**Fig. 11** Electron contribution to the specific heat ($\gamma$) in $MgCNi_{3-x}Ru_x$

**Fig. 12** Superconducting critical temperature ($T_C$) as a function of $\gamma$ in $MgCNi_{3-x}Ru_x$

**Fig. 13** $\Delta M/\Delta H$ at 10K as a function $\gamma$ of $MgCNi_{3-x}Ru_x$ samples

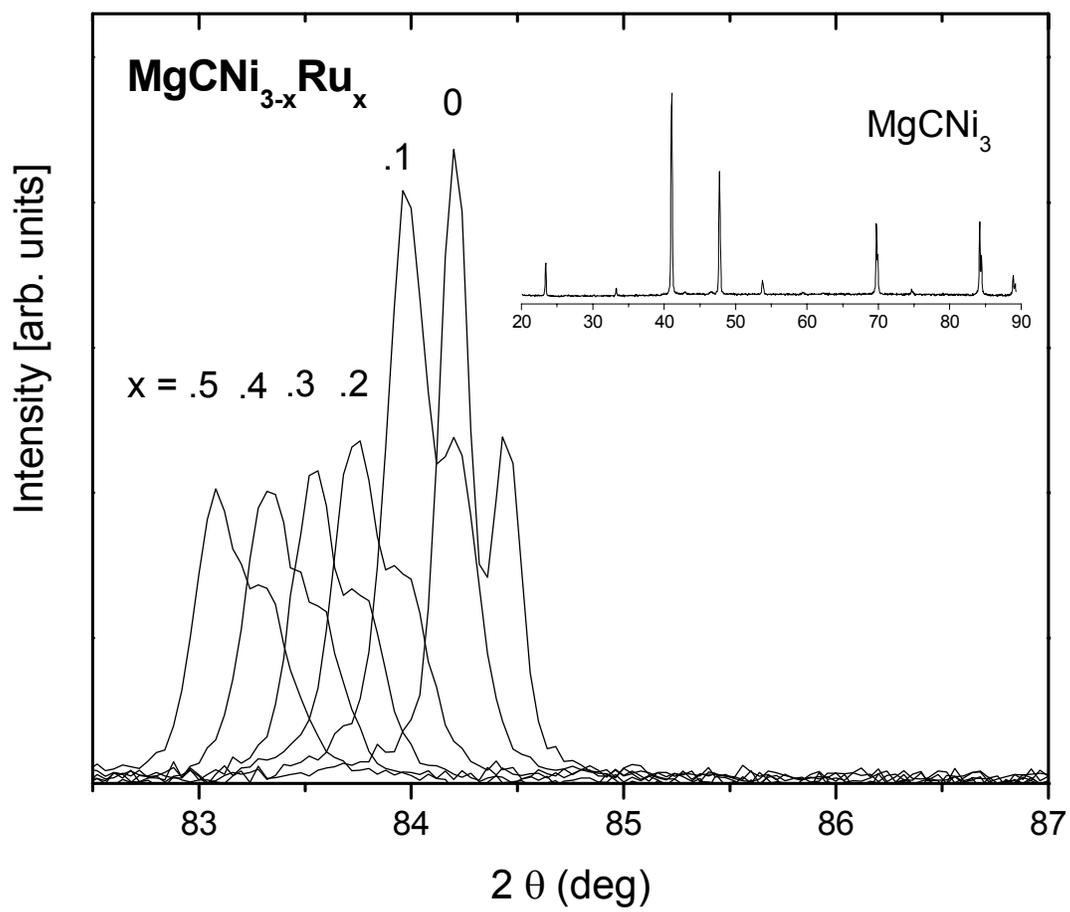

Fig.1

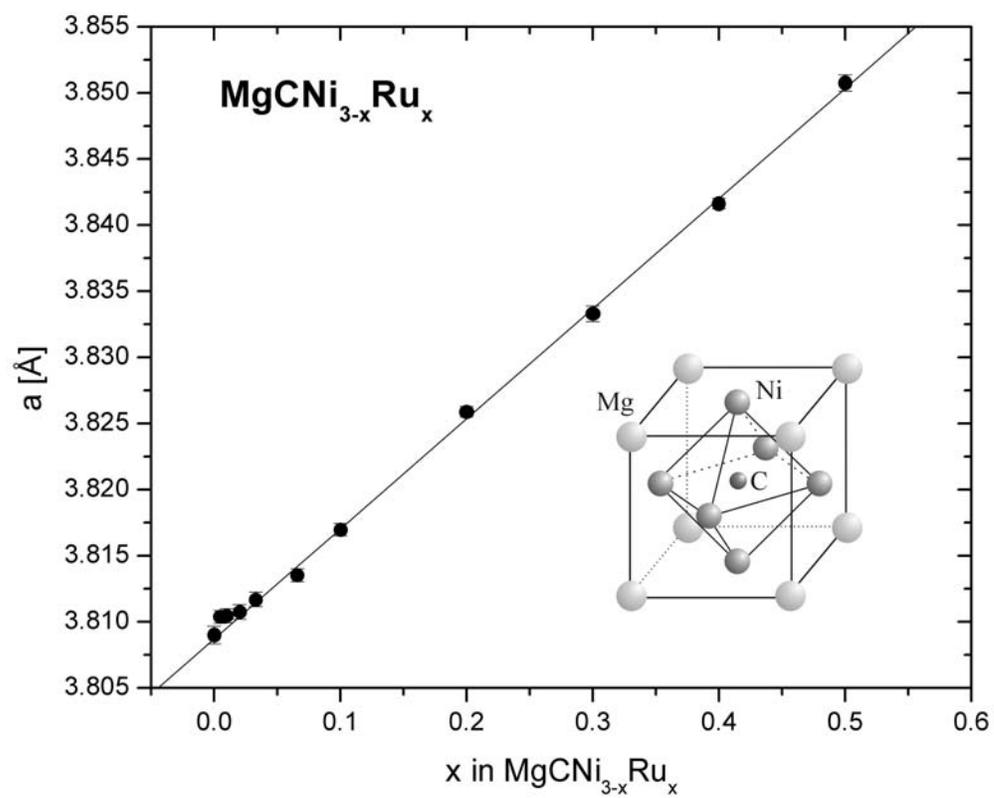

**Fig. 2**

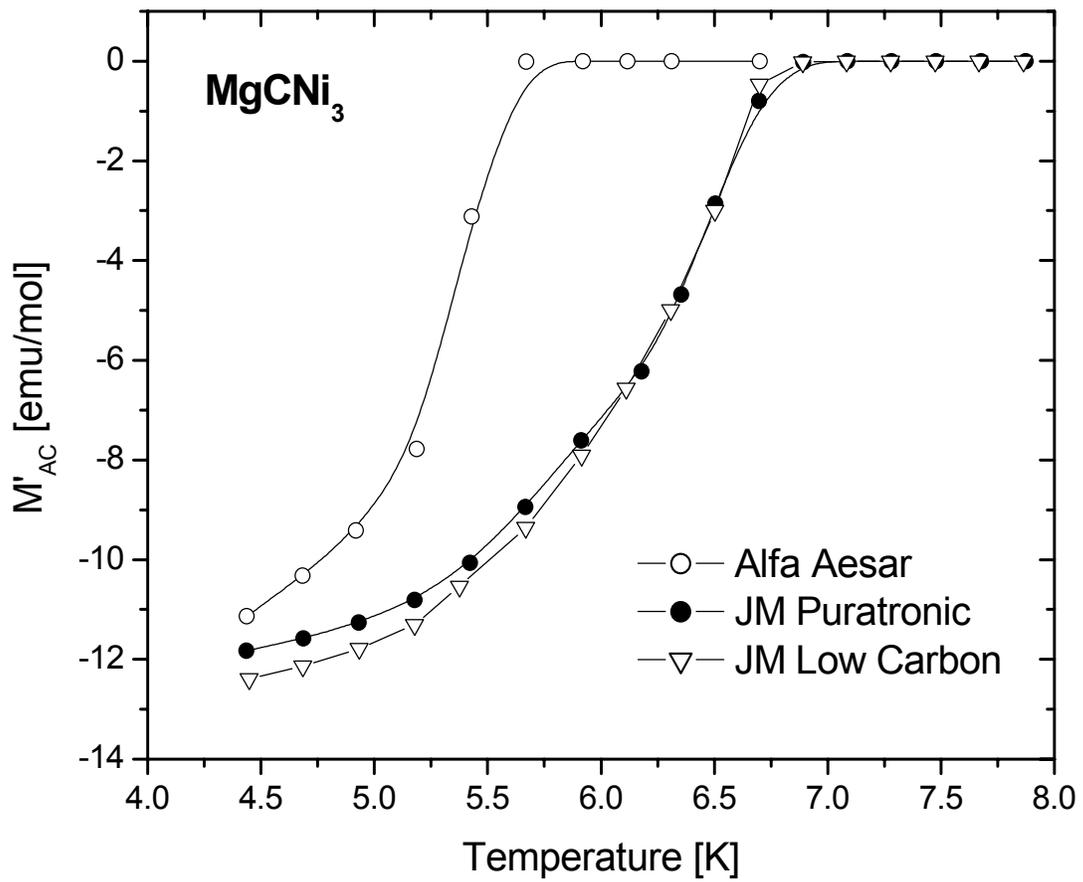

Fig. 3

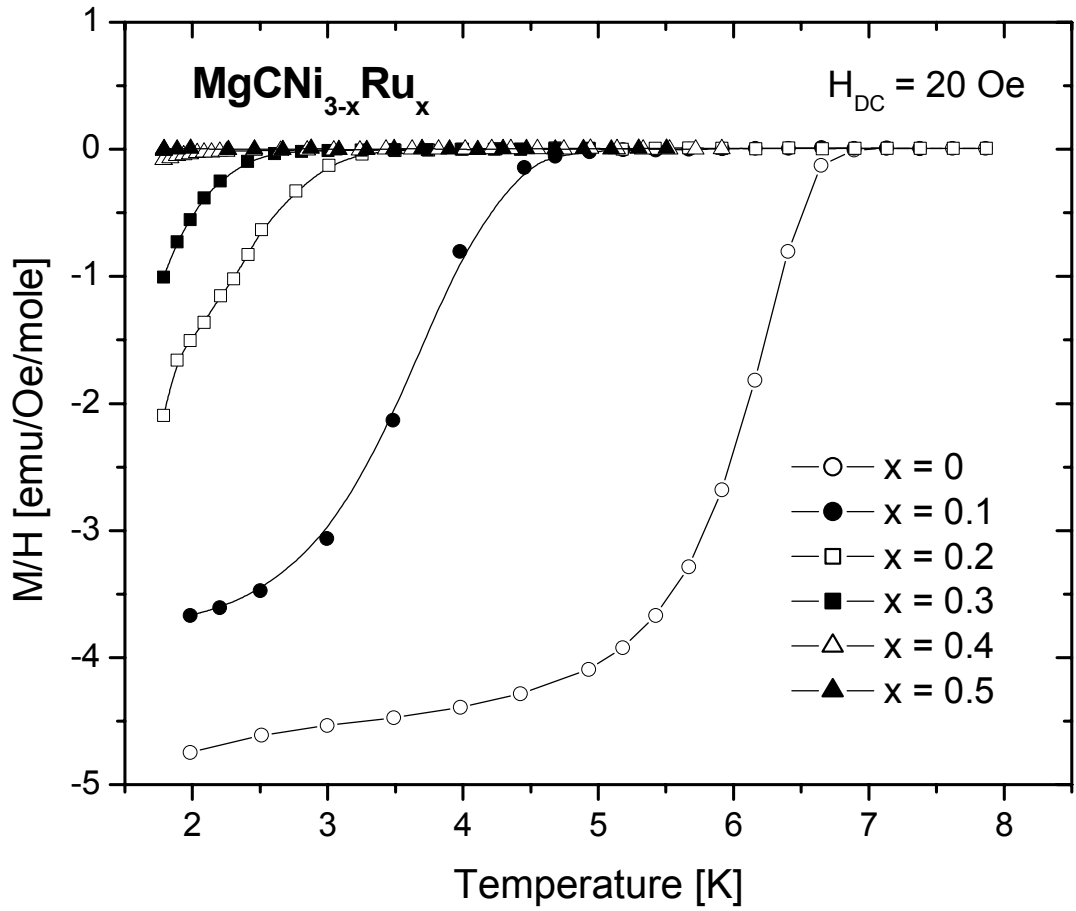

**Fig. 4**

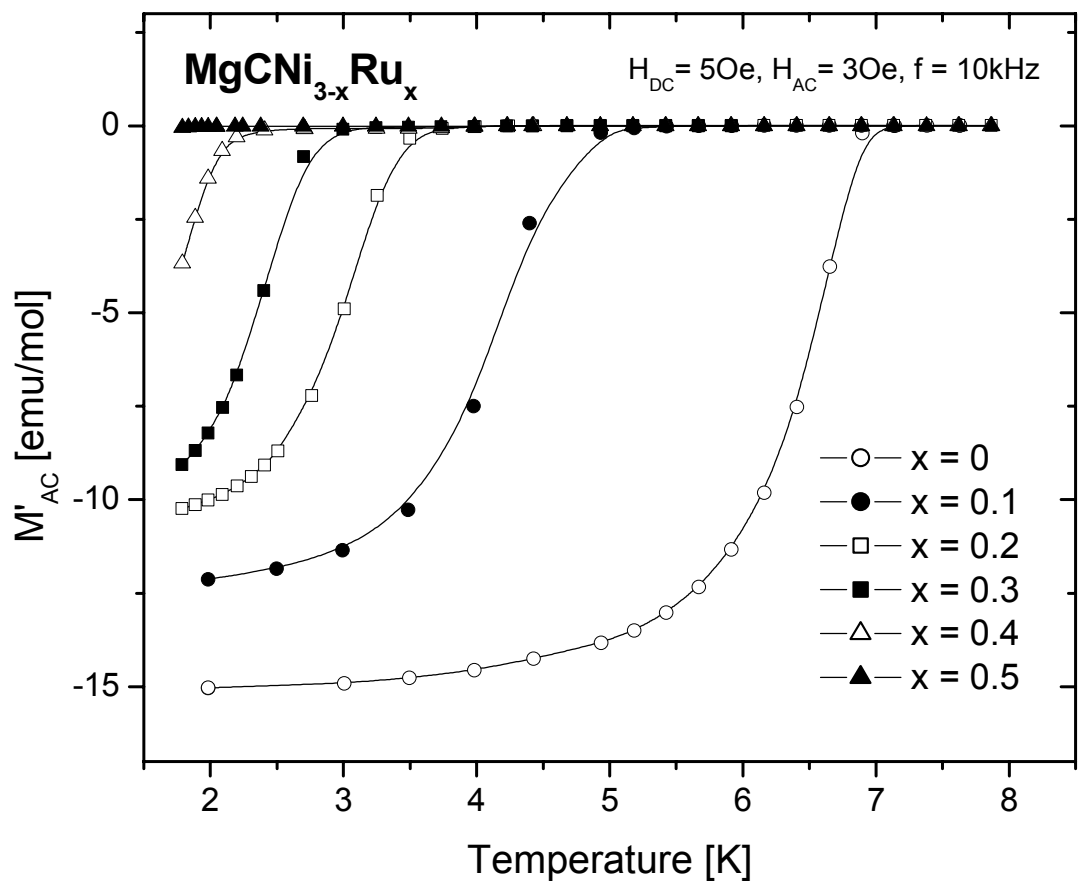

Fig. 5

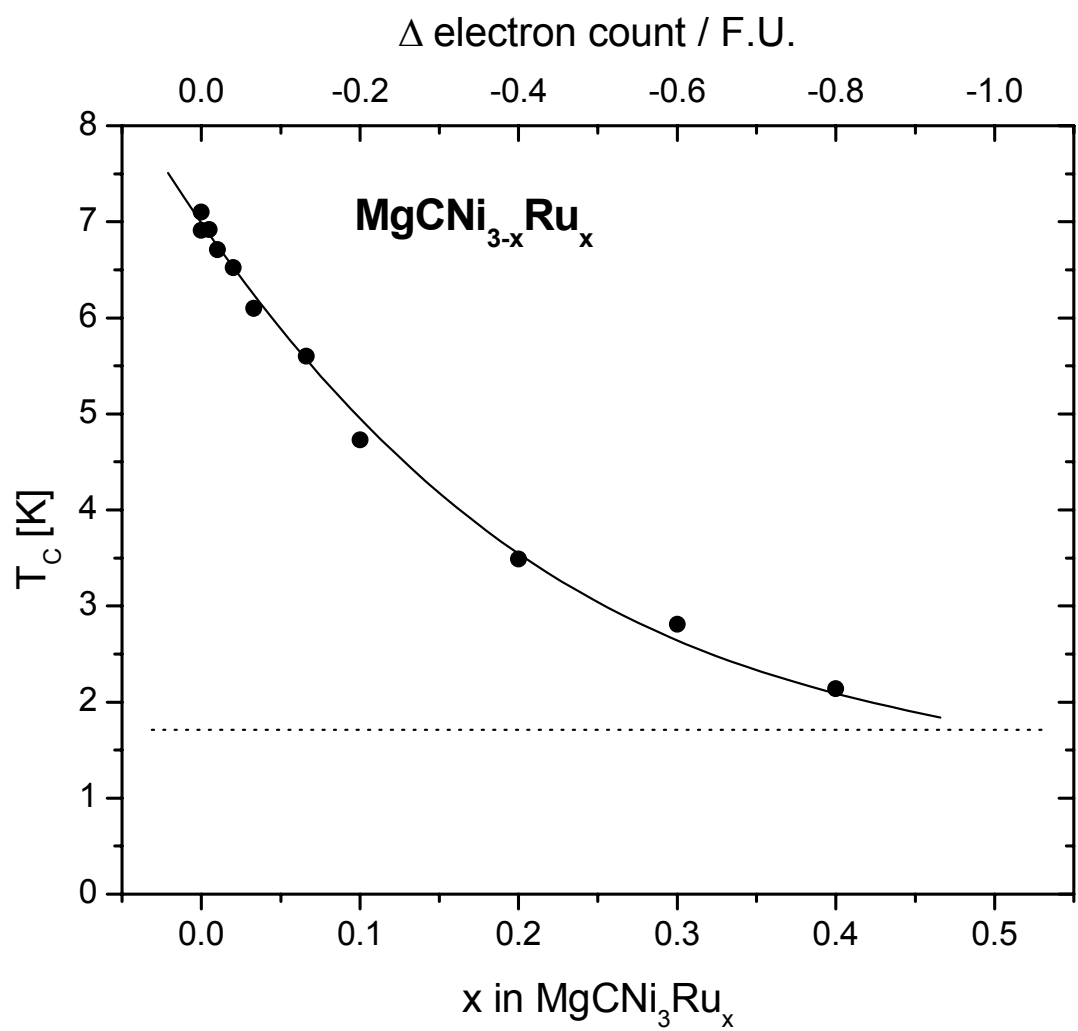

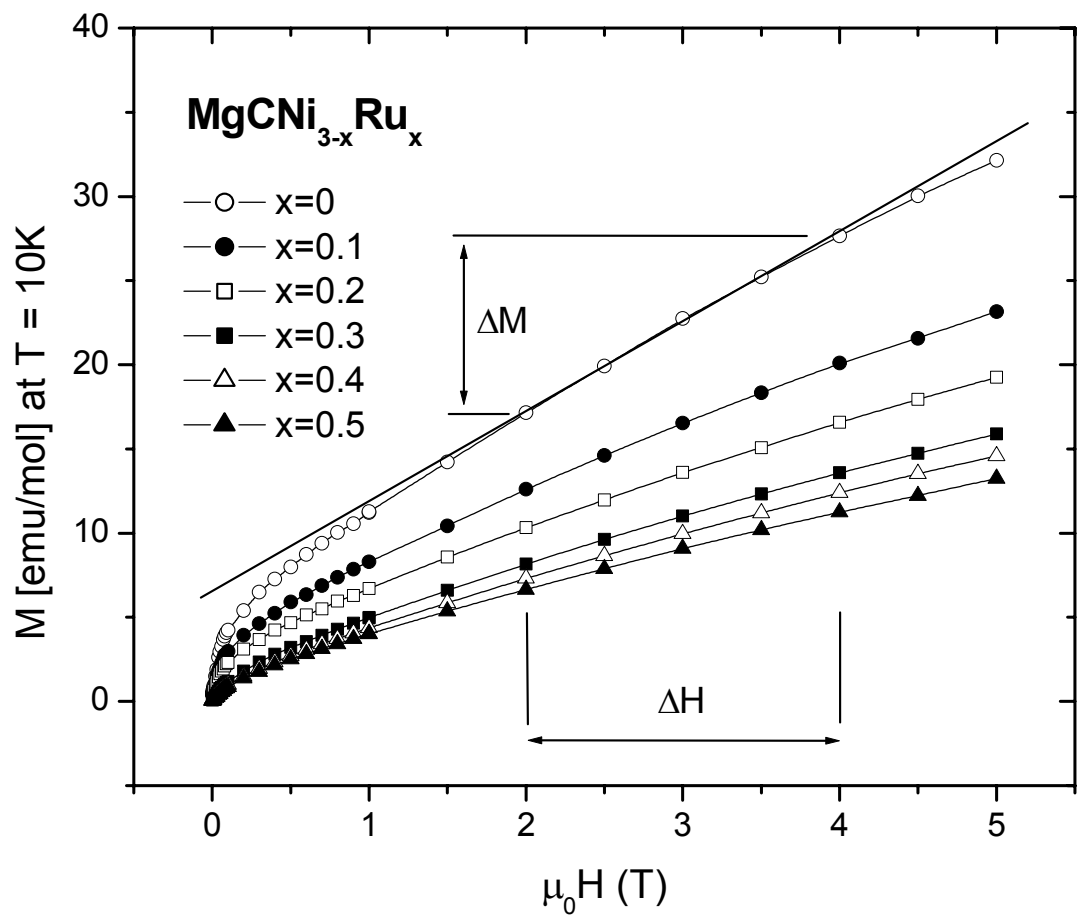

**MgCNi$_{3-x}$Ru$_x$**

- ○ x=0
- ● x=0.1
- □ x=0.2
- ■ x=0.3
- △ x=0.4
- ▲ x=0.5

M [emu/mol] at T = 10K

ΔM

ΔH

μ₀H (T)

**Fig 7.**

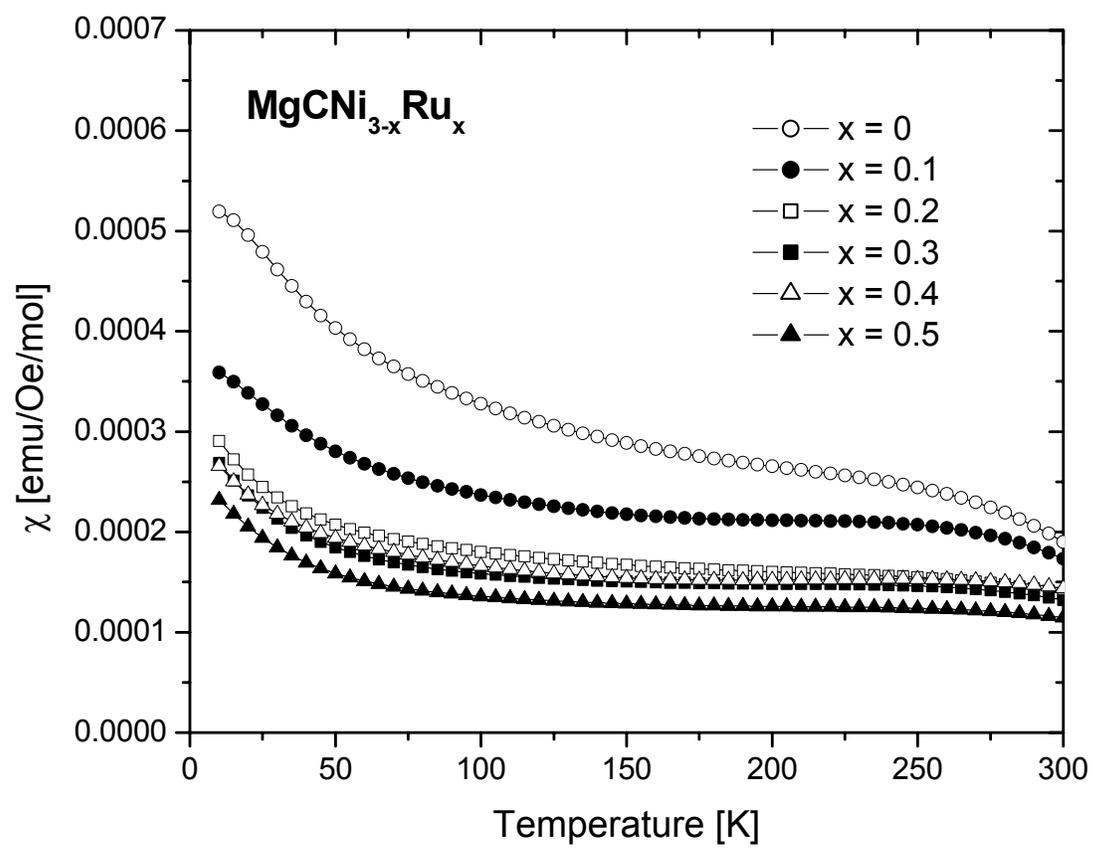

Fig. 8

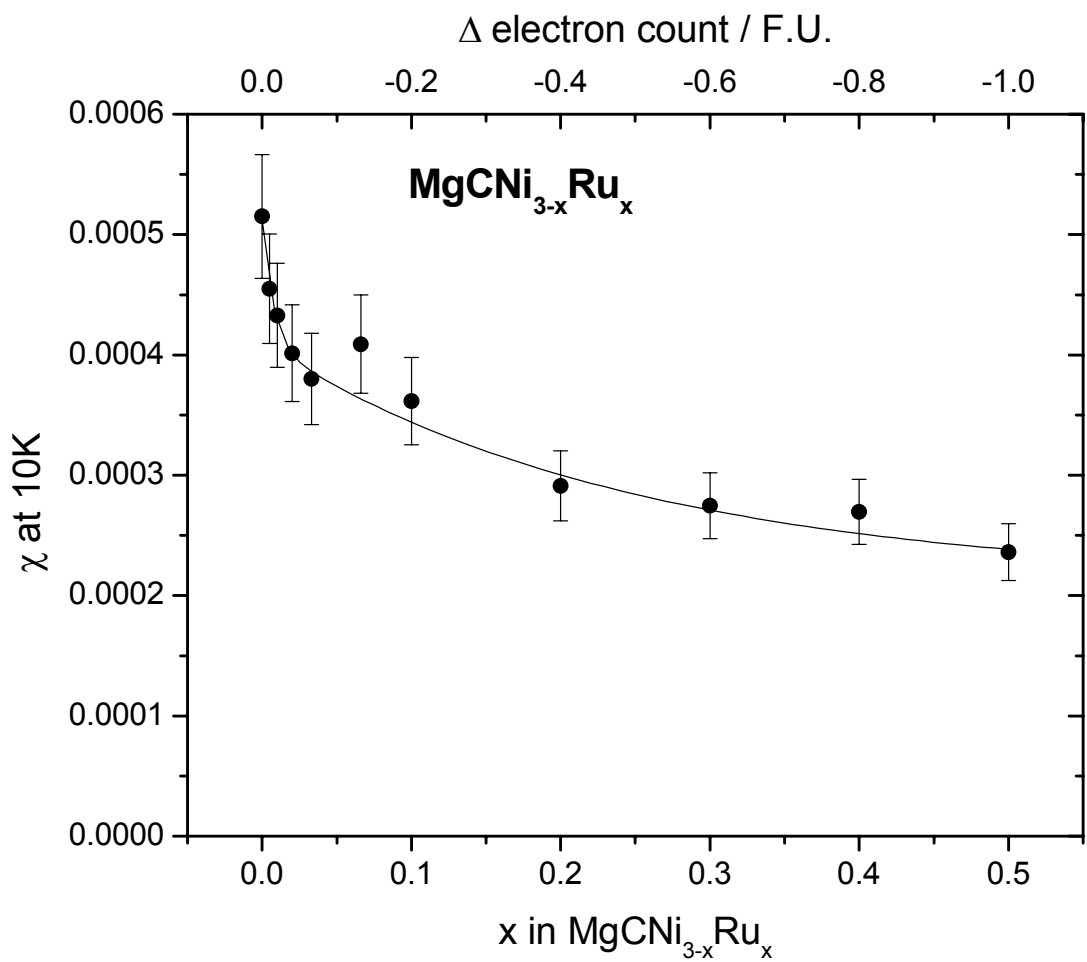

Fig. 9

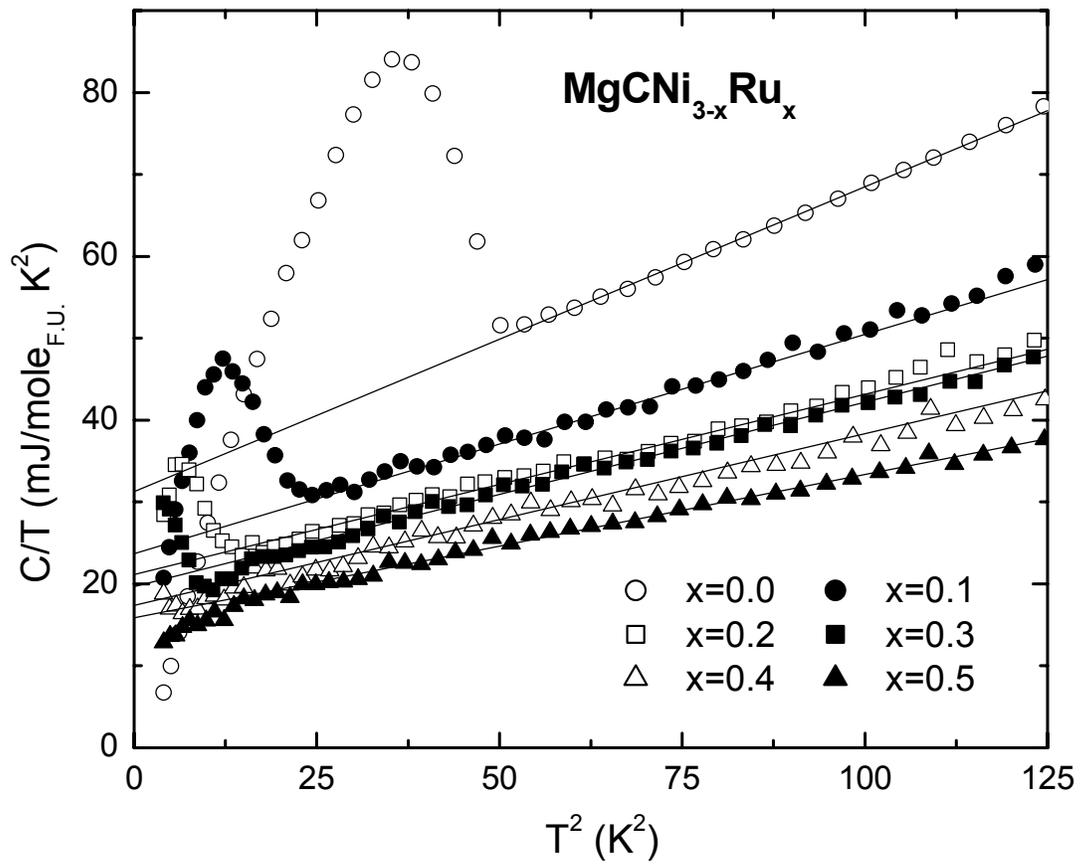

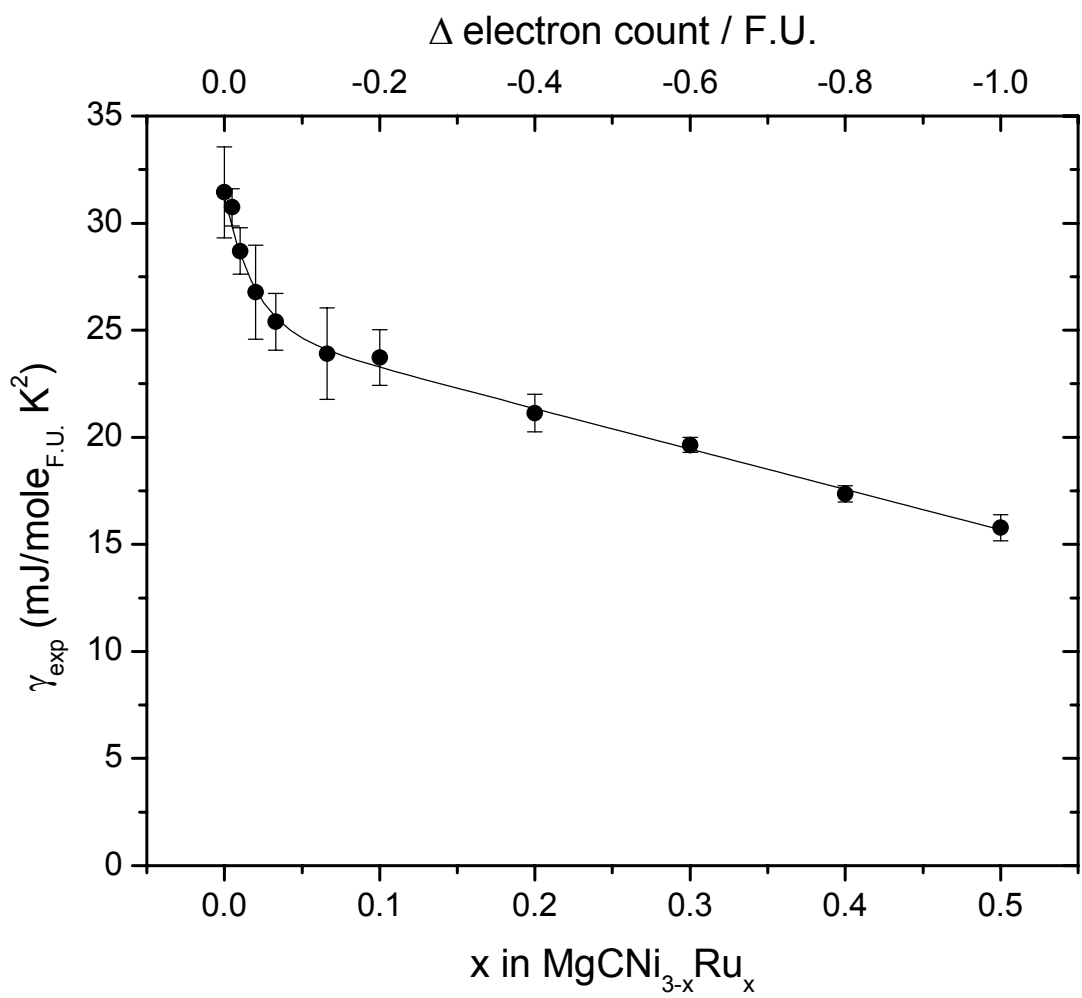

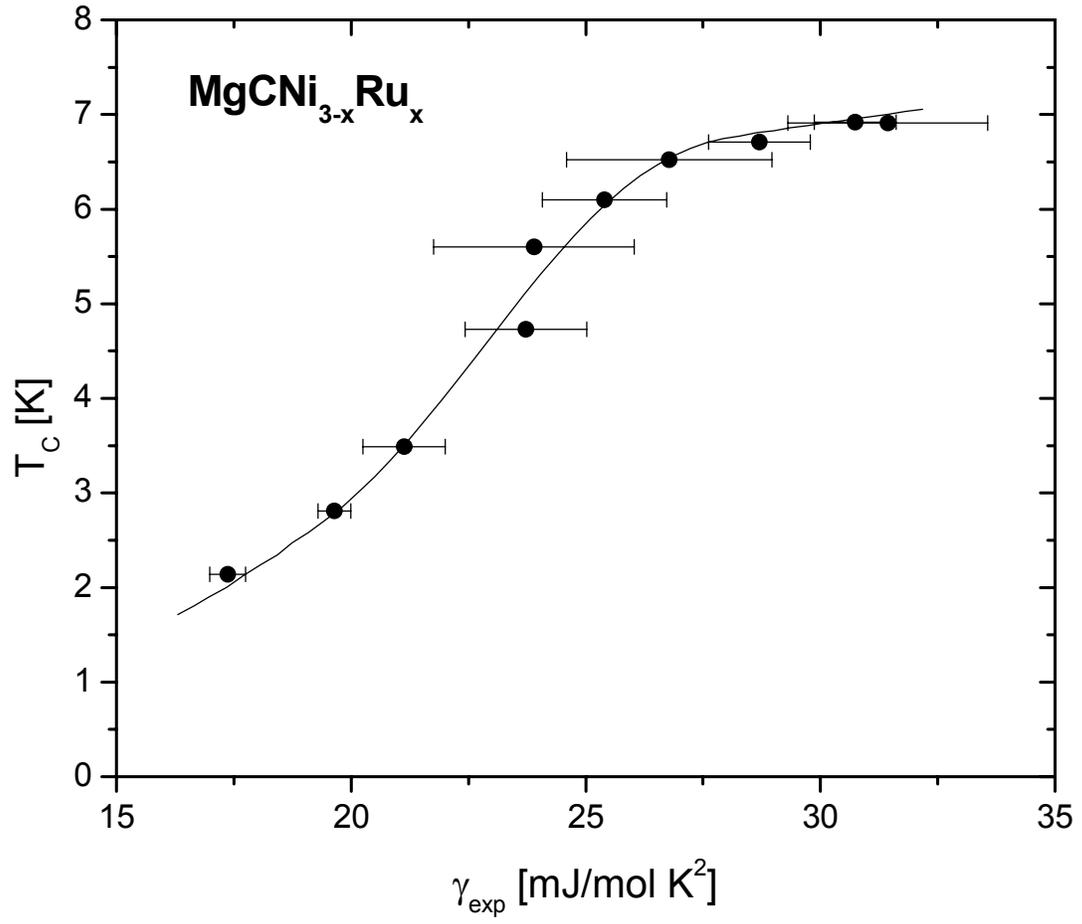

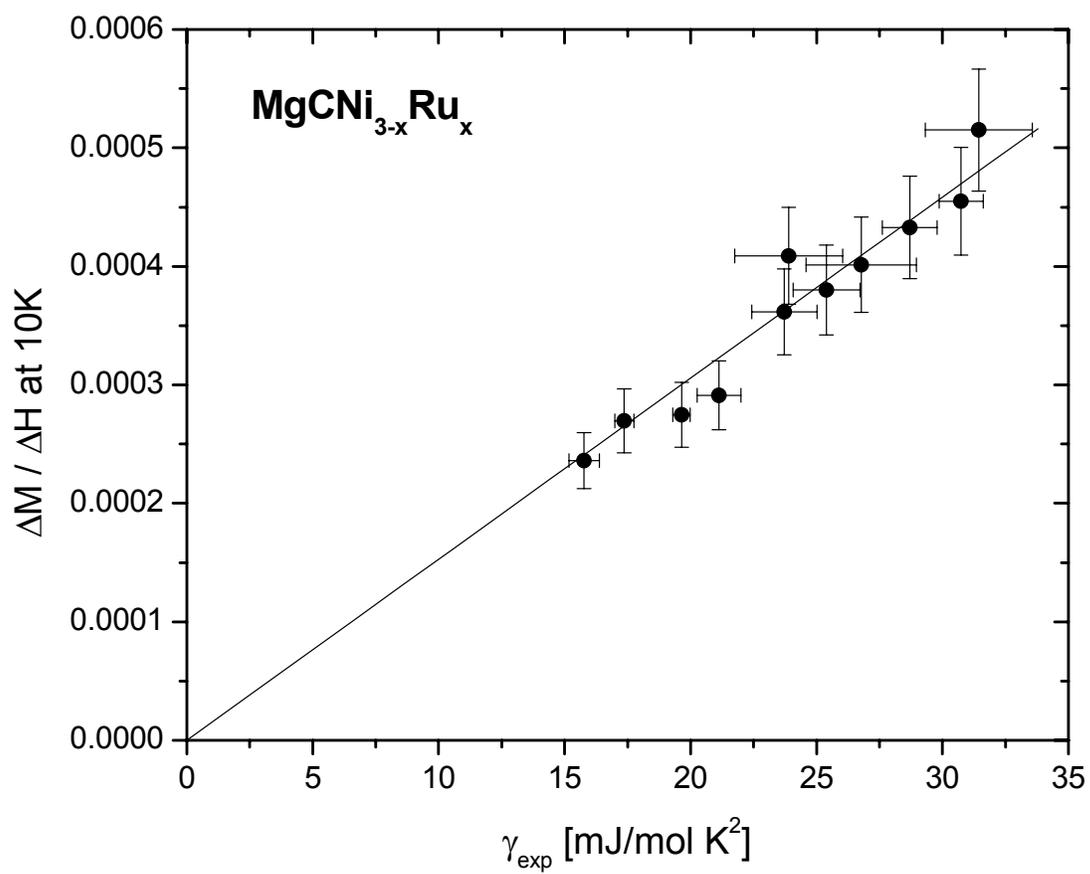

**Fig. 13**